\documentclass[prb,twocolumn,showpacs,floatfix]{revtex4}

\usepackage{graphicx}

\begin{document}

\title{Magnetic phenomena, spin-orbit effects, and Landauer conductance in 
Pt nanowire contacts}
\author{Alexander Smogunov,$^{1,2,3}$, Andrea Dal Corso,$^{2,4}$  
and Erio Tosatti$^{1,2,4}$}

\affiliation{$^1$International Centre for Theoretical Physics (ICTP), Strada Costiera 11, I-34014 Trieste, Italy}
\affiliation{$^2$INFM Democritos National Simulation Center, Via Beirut 2-4, I-34014 Trieste, Italy}
\affiliation{$^3$Voronezh State University, University Sq. 1, Voronezh, Russia}
\affiliation{$^4$International School for Advanced Studies (SISSA), Via Beirut 2-4, I-34014 Trieste, Italy}

\date{\today}

\begin{abstract} 
Platinum monatomic nanowires were predicted to spontaneously develop 
magnetism, involving a sizable orbital moment via spin orbit coupling, and
a colossal magnetic anisotropy. We present here a fully-relativistic 
(spin-orbit coupling included) pseudo-potential density functional 
calculation of electronic and magnetic properties, and of
Landauer ballistic conductance of Pt model nanocontacts consisting of 
short nanowire segments suspended between Pt leads or tips, reprented by
bulk planes. Even if short, and despite the nonmagnetic Pt leads, 
the nanocontact is found to be locally magnetic with magnetization 
strictly parallel to its axis. Especially under strain, the energy barrier 
to flip the overall spin direction is predicted to be tens of meV high, 
and thus the corresponding blocking temperatures large, suggesting the use
of static Landauer ballistic electrical conductance calculations. We carry out
such calculations, to find that inclusion of spin-orbit coupling and of magnetism 
lowers the ballistic conductance by about $15\div20$\% relative
to the nonmagnetic case, yielding $ G\sim 2 G_0$ ($G_0=2e^2/h$), in good agreement 
with break junction results. The spin filtering properties of this highly 
unusual spontaneously magnetic nanocontact are also analysed. 
\end{abstract}


\maketitle

\section{Introduction}
Metallic nanocontacts as thin as one atom can be fabricated, imaged and studied by
means of several experimental techniques including tip based instruments, 
transmission electron microscopy, and mechanically controllable break junctions.
In the latter, metal nanocontacts often break at the stage of single-atom contact;
in some cases in addition, short one-dimensional chains of atoms can be obtained 
prior to breaking.~\cite{ohnishi,yanson} It was shown experimentally\cite{smit1} 
and theoretically\cite{bahn,tosatti2005} that the formation of such atomic chains 
(monatomic wires) is favored in heavy $5d$ metals such as Ir, Pt, and Au. In Au, 
short suspended monatomic wire segments were first imaged experimentally by 
transmission electron  microscopy~\cite{ohnishi,rodrigues2001,kizuka} and their conductance, 
close to unity, characterized in break junctions.~\cite{agrait} 
Pt monatomic suspended nanowires were also characterized with similar methods.\cite{smit1,rodrigues2003}

Due to the nanometric dimensions, electron transport in atomic-sized nanocontacts
is essentially ballistic. For a nonmagnetic contact, or for a statically
magnetized one, ballistic conductance is given in the linear response limit
by the Landauer-B\"uttiker transmittance, $G=(1/2)G_0 \sum_i T_i(E_F)$, where 
$G_0=2e^2/h$ is the conductance quantum and $T_i(E_F)$ is the transmission 
probability at the Fermi energy for the conductance channel $i$.
In a nanocontact with a single atom cross section 
the number of conductance channels is controlled by the atom valency.
The simplest case of monovalent metals such as Au, Ag, and Cu 
presents just two $s$-like channels (one per spin) with almost perfect 
transmission so that the conductance is close to $ G_0$. 
In transition metals with partially occupied $d$ orbitals 
(e.g., Pt, Pd, Ni, Co etc.) in addition to two $s$ channels there are 
several $d$ channels also contributing to electron transport. This leads  
to a rather broad first peak centered well above $G_0$ in the conductance histograms. 
Here we shall be concerned with the case of Pt, of special interest 
owing to its proximity to magnetism, and to its known ability to form 
nanowires.\cite{smit1}

In Pt, a variety of values for the position of the first conductance peak 
have been reported, ranging from $0.5~G_0$ to $2.5~G_0$.
~\cite{smit1,olesen,costa-kramer,rodrigues2003,krans,yanson1,nielsen,smit2,untiedt} 
The origin of the subpeaks at $0.5~G_0$ and $1~G_0$ was later attributed
to the presence of gas molecules\cite{smit2,untiedt,garcia,strange} possibly
forming bridges between the electrodes just before contact breaking. 
For clean Pt nanocontacts data generally show a broad conductance 
histogram first peak centered between $1.5~G_0$ and $2~G_0$.
~\cite{smit1,krans,yanson1,nielsen,smit2,untiedt}
There have been to date a variety of theoretical studies of electron transport in Pt 
nanocontacts.\cite{garcia,garcia-suarez,strange,delavega,fernandez,pauly}
Recent calculations\cite{garcia-suarez,strange} based on density 
functional theory and on the Landauer-B\"uttiker formulation
reported conductances close to $2~G_0$ for single-atom contacts and 
for straight monatomic wires and lower values (down to $1.5~G_0$) for 
zigzag nanowires.

However, approaches used so far mostly restricted to the scalar 
relativistic level where spin-orbit coupling (SOC) is absent. 
Furthermore, only in one case magnetic effects have been considered 
in the calculation of the ballistic conductance.~\cite{fernandez}
Actually, Pt is a Stoner enhanced, non magnetic material in bulk; but 
the tendency towards magnetism is expected to get stronger in monatomic 
chains,\cite{delin} for two reasons. The first is that the narrowing of 
$d$ bands caused by the lower atomic coordination acts to enhance the 
electronic density of states (DOS) at the Fermi energy to approach and
eventually to exceed the Stoner instability limit. 
This effect is especially pronounced in one-dimensional Pt 
monatomic wires where van Hove band edge singularities fall near the
Fermi level.\cite{delin,colossal} The second reason stems from SOC, which 
reinforces spin magnetism with an accompanying orbital moment.
In a Pt monatomic wire, SOC acts to stabilize magnetism, raising 
a sizeable parallel orbital moment to accompany a nonzero spin moment, even at 
the equilibrium interatomic distance. In density functional calculations,
a spin-orbit split electronic band edge is pushed up closer to the Fermi level 
driving the infinite monatomic nanowire to a ferromagnetic ground state.~\cite{delin, colossal} 

Our interest here is on the conductance of a monatomic Pt nanocontact, on the 
role of SOC, and on the effect of local magnetization if present. Because
magnetization in a (``zero-dimensional'') nanocontact will generally fluctuate, 
a static calculation such as that given here, and the use of the Landauer-B\"uttiker 
approximation to the conductance are not in principle adequate. Very close 
to zero voltage, dynamical phenomena such as Kondo anomalies may affect the 
conductance. In the present study we will leave this regime aside and 
defer these interesting dynamical phenomena to a later study. We will 
therefore restrict to the basic Landauer-B\"uttiker static level. This
should be adequate at small but finite voltages above possible
zero-bias anomalies. Our calculations will demonstrate chiefly the role 
of SOC and of local magnetism on the nanocontact ballistic conductance. 
They will also demonstrate that the recently discussed property of colossal 
magnetic anisotropy and of large energy barriers against flip of magnetization 
\cite{colossal} should also apply to monatomic nanocontacts, with the
interesting consequences of large ``blocking'' temperatures.  
 
We recently presented an approach suitable for calculating the Landauer-B\"uttiker 
conductance based on a generalization of the method of Choi and Ihm~\cite{choi} to the case 
of fully relativistic (FR) ultrasoft pseudo-potentials (US-PPs).~\cite{cond_soc} 
In Ref.~\onlinecite{mosca-conte} we showed that these PPs can reproduce 
accurate electronic band structures of fcc-Pt and fcc-Au calculated by solving 
a four-component Dirac equation, within a scheme based on two component
spinors. In this paper, we will use these FR US-PPs to study theoretically 
short monatomic Pt wires suspended between two bulk Pt leads and
address several important issues. First, does the suspended nanowire segment
remain locally magnetic even when just a few atoms long, and despite being
attached to bulk nonmagnetic leads? And, since we will find that this is 
the case, what is the role of the wire's length and strain on the local nanowire 
magnetism? Further, in presence of SOC, does the nanowire contact still exhibit 
giant or colossal magnetic anisotropy and the large energy barrier to magnetization 
reversal characteristic of the infinite nanowire? What is the joint effect of 
SOC and of magnetism on the ballistic conductance of Pt model nanowire contacts? 
And finally, how spin selective will the conductance be, and what kind of spin 
filtering would the static local moment magnetization exert on the current? 
The answer to these questions, so far unknown, should provide new basic
information on the physics of these unconventional nanocontacts, with an 
underlying potential and interesting connection between magnetism and conductance.

\section{Infinite monatomic wire}
Before focusing on Pt nanowire contacts we will begin by presenting some results for 
the idealized case of an infinite tipless monatomic wire, which reproduce
those recently published by us and by others.\cite{delin,colossal,garcia-suarez,fernandez}
Electronic structure calculations were carried out within the density
functional theory, using the standard plane-wave {\tt PWscf} code
of the QUANTUM-ESPRESSO package.\cite{pwscf}
The local-spin-density approximation (LSDA) in the form introduced by 
Perdew-Zunger\cite{perdew} is used for the exchange-correlation energy.
We performed both scalar-relativistic (SR) and fully-relativistic (FR) 
pseudo-potential calculations. 
In the first case, the Pt ions are described by US-PPs generated with 
all relativistic corrections except of the spin-orbit coupling. In the 
second case, the fully-relativistic US-PPs (with spin-orbit included) are 
used to simulate Pt ions and electron wave-functions are two-component 
spinors. The parameters of the PPs are given in Ref.~\onlinecite{mosca-conte}.  
The cutoff kinetic energies were 30 Ry and 300 Ry for the wave functions 
and for the charge density, respectively. Integration over the Brillouin 
zone (BZ) up to the Fermi energy was performed by using
320 one-dimensional k points with a smearing parameter of 
0.002 Ry.\cite{smearing} 

Both SR and FR total energy minimizations for an infinite Pt wire yield 
a zero strain nanowire equilibrium interatomic distance $d_0\approx$ 2.35 \AA. 
However, while the SR calculation predicts the unstrained Pt wire to be nonmagnetic,
the FR calculation yields for the same system a ferromagnetic ground state with the 
magnetization parallel to the wire, in agreement with earlier calculations of Delin 
{\it et al.}~\cite{delin} 
The calculated spin and orbital moments per atom at $d_0$ are 
$M_{S\parallel}=0.17~\mu_B$ and $M_{L\parallel}=0.22~\mu_B$, respectively. 
Furthermore, the magnetization magnitude vanishes if its direction 
is constrained to lie perpendicular to the wire. A magnetic state with transverse 
magnetization actually becomes stable with strain, but only at 
much larger interatomic distances (above 2.6 \AA), and even then at a much 
higher energy than the nanowire ground state with parallel magnetization.
Thus the equilibrium and low strain Pt nanowire displays a {\it colossal} 
magnetic anisotropy with the easy axis parallel to the wire.~\cite{colossal}
The (zero-temperature) infinite Pt nanowire is actually quite a strong magnet,
especially when strained. We gauged for example the strength of interatomic 
magnetic exchange by forcing a magnetization reversal and found an energy increase 
of about 85 meV at $d=$ 2.66 \AA. In a localized picture of magnetism, this
would be equivalent to a near neighbor intersite exchange magnitude of nearly
500 K.  

\begin{figure}
\hskip-3mm
\includegraphics[width=8.5cm]{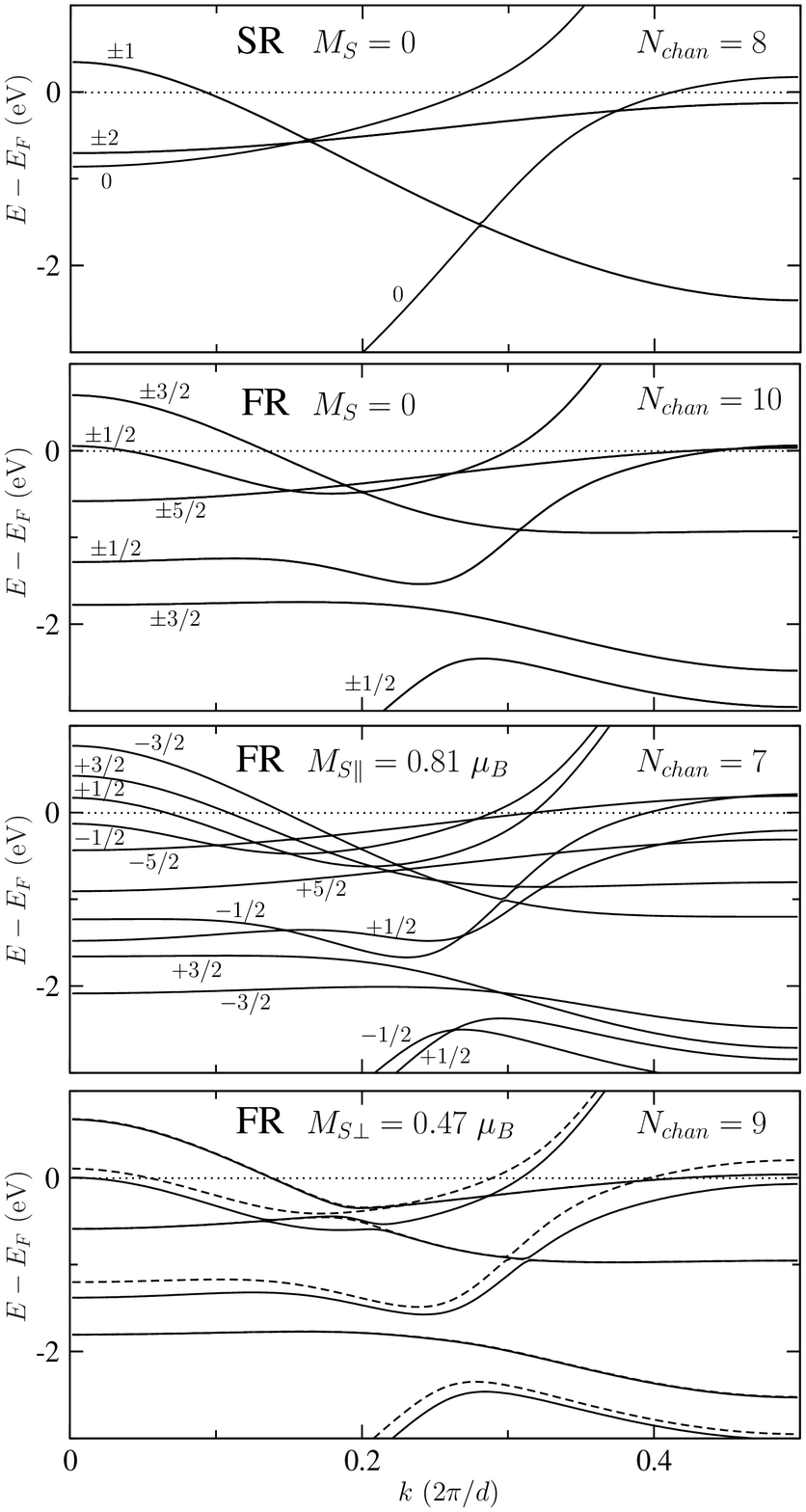}
\caption{\label{bands}
Electron band structure of the infinite monatomic Pt wire with interatomic distance of
2.66 \AA~ calculated with SR (upper panel) and FR (lower panels) pseudo-potentials.
For the FR, more realistic case, the bands for nonmagnetic as well as for 
both magnetic polarizations (magnetic moment parallel or perpendicular to the 
wire axis) are shown. Bands are labeled according to their symmetry (see text).
The number of bands crossing the Fermi level (number of conductance channels) 
is provided on each panel. The spin magnetic moment per atom is also given for 
magnetic states.
}
\end{figure}

Interatomic distances at break junction nanocontacts are generally 
under heavy strain. Measured break junction forces in fact generally lie 
above 1 nN.\cite{rubio} Especially at criogenic conditions where 
atomic mobility is largely frozen out, stress and strain are generally 
speaking not well characterized. At larger temperatures, and in conditions 
of quasi equilibrium, it can be argued that a finite spontaneous stress 
tension should be present even when the tips are kept at fixed distance 
without any pulling force.\cite{tosatti2001,tosatti2005} 
In our nanowire study we arbitrarily choose strains ranging 
from an interatomic Pt-Pt distance $d= d_0=2.35$~\AA~ (zero strain) to $d=2.66$~\AA~ 
which corresponds to a strained nanowire, with a tension of approximately 3.8~nN,
roughly in the range of spontaneous breaking
(we checked that a nanocontact formed by a three-atom segment
of such a strained nanowire is still stable, see next section).
At this interatomic distance the infinite wire is still nonmagnetic within the SR 
scheme, where SOC is absent, whereas in the more correct FR case which includes SOC, 
the wire is strongly ferromagnetic. The colossal magnetic anisotropy 
here reverts to a conventional giant anisotropy\cite{mokrousov,viret,autes} 
with an easy magnetization axis parallel to the wire and an extremely 
large anisotropy energy, $E_{\perp}-E_{\parallel}\approx 36$ meV/atom.

Qualitatively relevant to our subsequent reasoning on electron transport 
is the number of conductance channels $N_{chan}$, that is the number 
of bands crossing the Fermi level. In Fig.~\ref{bands} we show the electron 
band structure of the infinite Pt monatomic nanowire calculated in the SR 
and FR schemes. In the latter case SOC splits the bands in the (artificially 
forced) nonmagnetic state, and magnetism further splits them corresponding 
to majority and minority polarizations. The ballistic conductance 
channels are clearly seen to change from one case to another. 

Although unrealistic in practice, the infinite nanowire is important
because it lends itself to understand symmetry aspects. We classify the 
electronic bands as follows. In the SR case the Hamiltonian commutes with 
$\hat{L}_z$, the projection of the orbital angular momentum on the wire axis 
(the $z$ axis) so that electron states can be labeled by its eigenvalues 
$m=0,\pm1,\pm2,$ etc. (in units of $\hbar$).
Bands with $\pm m$ are degenerate due to the mirror symmetry with respect
to a plane such as $xy$ containing the wire axis.
This classification corresponds to one-dimensional ($m=0$) and two-dimensional
($m=\pm1,\pm2$, etc.) irreducible representations of the symmetry group $C_{\infty v}$. 
Moreover, all bands possess additional degeneracy due to 
spin, since the spin and orbital degrees of freedom do not mix 
in the SR case. In the FR case, the spin and orbital degrees of freedom are 
coupled through SOC and the bands are now labeled by half-integer eigenvalues 
$m_j=\pm1/2,\pm3/2,\pm5/2,$ etc. of the operator $\hat{J}_z$, the $z$-component of the total 
angular momentum.
In the nonmagnetic state, again due to the mirror symmetry in 
the $xy$ plane (and also due to time reversal symmetry in conjunction with 
inversion symmetry) the bands with $\pm m_j$ (an infinite number of two-dimensional 
irreducible representations of the double group of $C_{\infty v}$) are degenerate. 
The two-fold degeneracy of these nonmagnetic states is lifted when the wire 
is magnetized in the parallel direction since reflection (as well as time reversal) 
is no longer a symmetry operation. Magnetization, an axial vector, reverses 
its direction under both spatial reflection in the $xy$ plane and under time reversal, 
but not under inversion. This lifting of degeneracy corresponds to a reduction 
of the symmetry group from $C_{\infty v}$ to $C_{\infty}$, the latter with
one-dimensional irreducible representations only. In the state with transverse 
magnetization (lower panel), the only remaining symmetry is a reflection 
through the plane containing the wire axis and perpendicular to the magnetization.
The symmetry double group is $C_s^D$ with two one-dimensional 
irreducible representations  $\Gamma_3$ and $\Gamma_4$. Hence, 
all bands separate here into two groups transforming according to $\Gamma_3$ or $\Gamma_4$ 
shown by solid and dashed lines, respectively. 

We find that the number of conductance channels is modified both 
by spin-orbit interactions and by magnetism. Comparing the two 
upper panels we see for example that the spin-orbit interaction splits 
four $m=\pm2$ bands into the pair of two-fold degenerate bands with 
$m_j=\pm3/2$ and $m_j=\pm5/2$. With SOC but still no magnetism the bands 
with $m_j=\pm5/2$ are pushed up in energy to cross now the Fermi 
level so that the number of conductance channels is increased from 8 to 10. 
We note that there are three two-fold degenerate bands 
(one with $m_j=\pm5/2$ and two with $m_j=\pm1/2$) with edges very close 
to the Fermi energy, which results in an exceedingly large DOS at the $E_F$. 
Upon onset of parallel ferromagnetism, this band edge near $E_F$ is split
by magnetic exchange, one majority band now completely full, and
a minority band further emptied. Nanowire magnetization lowers the DOS at $E_F$,
corresponding to a sort of band Jahn Teller effect.\cite{colossal}  
With parallel magnetization three conductance channels altogether drop 
out compared to the nonmagnetic state, whence now $N_{chan}=7$. 
The number of conductance channels increases by two, from 7 to 9, 
when the magnetization is rotated from parallel to perpendicular
to the wire axis. The corresponding change in conductance with the 
magnetization direction, also termed ``anisotropic magnetoresistance'' (AMR)
can be characterized by the ratio $(G_\perp-G_\parallel)/G_\parallel$.
Our infinite nanowire calculations thus foreshadow a large AMR
even in real Pt contacts, provided they will develop a spontaneous 
magnetic moment, and provided that the moment could be rotated with
practically attainable external magnetic fields. We will see later that 
while the former is confirmed, the required field to rotate magnetization 
in a Pt contact may generally be very large and hard to reach in practice. 
In the present academic case of the infinite nanowire, 
AMR$=(N_{chan,\perp}-N_{chan,\parallel})/N_{chan,\parallel}\approx 28{\%}$.

\section{Monatomic nanowire contact: magnetic properties}

\begin{figure}
\hskip-3mm
\includegraphics[width=8.5cm]{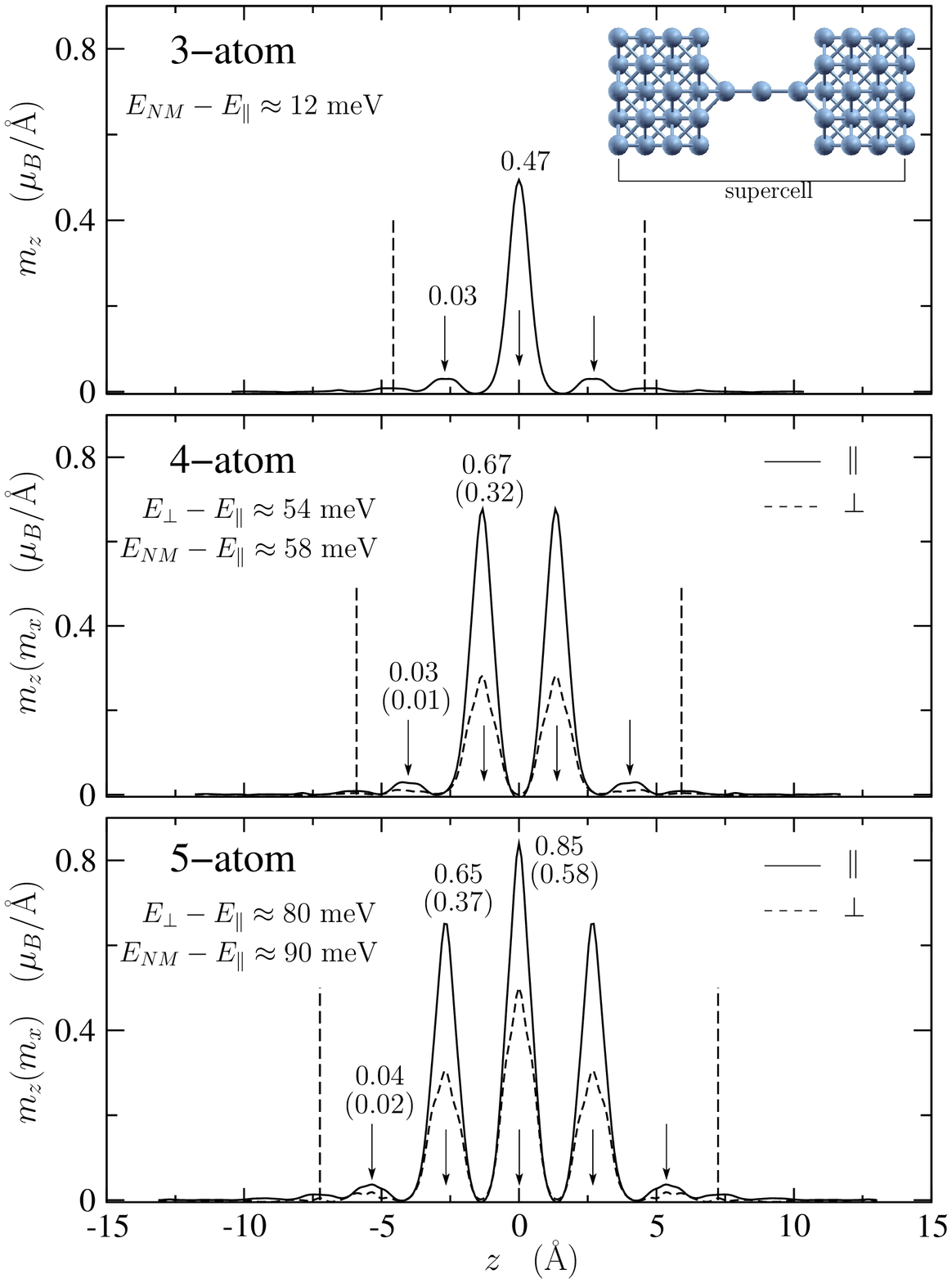}
\caption{\label{mag_3a4a5a}
Planar average (in the $xy$ plane) of the spin magnetization as a function of $z$ 
for three-, four-, and five-atom Pt nanowire contacts calculated with FR pseudo-potentials. Magnetic states 
with parallel and perpendicular polarization in the four- and 
five-atom case are shown by solid and dashed lines, respectively. Magnetic moments 
(in Bohr magnetons) within a sphere with radius of 2.5~\AA~are shown for all nanowire atoms.
Vertical dashed lines show the positions of the bulk lead surfaces; the 
positions of nanowire atoms, with a mutual spacing of $d=2.66$\AA, are indicated by arrows.
}
\end{figure}

Next to the academic infinite nanowire, we considered model Pt contacts, made 
up of a short $N-$atom linear chain segment suspended in vacuum between two semi-infinite 
ideal Pt bulks. In line with real break junction data,\cite{smit1} we considered 
contacts with $N=$ 3, 4, and 5 atoms in the chain. The contacts consist of the supercell 
depicted for $N=$ 3 in the inset of Fig.~\ref{mag_3a4a5a}. The
nanowire segment is attached at ``hollow" sites of two mirror symmetrical
Pt(001) surfaces. For calculations, a ($2\sqrt{2}\times2\sqrt{2}$) supercell periodicity is 
employed in the $xy$ plane, perpendicular to the nanowire. We checked that
this spacing is sufficient to keep the periodically repeated wires 
enough apart from one another to make their mutual influence irrelevant. 
The bulk ``leads'' are simulated by a planar slab consisting of seven atomic 
$(001)$ crystalline planes (8 atoms per plane), sufficient to reproduce a 
bulk-like potential in the middle of the slab. Periodic boundary conditions 
are assumed in all three directions for electronic structure calculations. 
Since the supercell is very large along [001] (the $z$ axis),
the BZ in this direction is sampled only at $k_z$ = 0 while in 
the ($k_x, k_y$) plane perpendicular to the wire convergence demanded 
instead ten two-dimensional (2D) special {\bf k} points. This level of BZ sampling 
was checked for accuracy and found to be sufficient for obtaining a converged 
self-consistent potential needed for subsequent transmission calculations.
The energy smearing parameter near $E_F$ was chosen to be $0.01$~Ry.

Atoms of the seven-layer slab representing the bulk leads are 
located at their ideal bulk fcc positions (with interatomic spacing of 2.77 \AA).
The interatomic distance in the nanowire for all the contacts was 
kept at the same strained value earlier chosen for the infinite
nanowire, namely $d=$2.66 \AA~. The distance between the (001) surface planes and 
the last contact nanowire atoms was 1.91 \AA, a value obtained by 
prior SR calculations of the three-atom contact and optimization moving 
the contact nanowire atoms until forces acting on them
vanished (we note that in doing so we were also changing the length of 
the supercell in the $z$ direction). We checked that this geometry 
corresponds to a strained but still unbroken chain nanocontact.

The main result is that for all chain lengths $N$ the Pt nanocontacts
develop local magnetism. In Fig.~\ref{mag_3a4a5a} we plot the planar 
average (in the $xy$ plane) of the spin magnetization as a function of $z$ 
for the nanowire contacts of different lengths calculated with FR
pseudo-potentials, fully including SOC. We discuss first the three-atom nanocontact 
(upper panel). Starting with the three chain atoms magnetized in the parallel direction 
(as in the ground state of the infinite wire) we found that the nanocontact 
remains locally magnetic despite being in strong contact with 
nonmagnetic bulk Pt leads. The local magnetization is maximum at the 
central nanowire atom, and is significantly suppressed at the contact 
atoms touching the nonmagnetic Pt leads. The central nanowire Pt atom 
has a spin moment of 0.47 $\mu_B$, to be compared with the corresponding value 
for the infinite wire (0.81 $\mu_B$). The {\em total} nanocontact spin magnetic 
moment $M_{S\parallel}^{tot}$  was 0.56 $\mu_B$. Of this, about 0.03 $\mu_B$ is due to evanescent magnetization tails propagated from the magnetic Pt chain into the two bulk leads. 
Starting next with an initial magnetization transverse to the chain, we 
found that at $N=$ 3 this polarization does not survive, converging to a 
nonmagnetic nanocontact state. The infinite nanowire transverse magnetization,
already weak from the beginning, is in this case suppressed by the nonmagnetic 
bulk Pt leads so that such contact displays the property of colossal magnetic 
anisotropy.\cite{colossal} 

The influence of the nonmagnetic leads is expected to weaken for increasing
nanocontact chain length $N$. The results for $N=4$ and 5 
are shown in Fig.~\ref{mag_3a4a5a} on two lower panels. In these longer nanocontacts both 
magnetic states with parallel and transverse magnetizations are 
sustained. The corresponding magnetization curves are shown 
by solid and dashed lines for the parallel and transverse case, 
respectively, showing that magnetism indeed becomes more robust with 
increasing nanocontact chain length. For example, in the ground state 
(magnetization parallel to the wire) the magnetic 
moment of the chain central atom grows from 0.47 $\mu_B$ ($N=3$) 
to 0.67 $\mu_B$ ($N=4$), and further to 0.85 $\mu_B$ ($N=5$). The latter is 
in fact slightly larger than the magnetic moment per atom in the infinite 
wire (0.81 $\mu_B$), a slight overshoot possibly related to surface induced 
interference effects.
The total nanocontact spin magnetic moment here was $M_{S\parallel}^{tot}$=1.44 $\mu_B$ and 
$M_{S\parallel}^{tot}$= 2.29 $\mu_B$ for four- and five-atoms nanocontacts, respectively. 
The amount of spin moment spread in the bulk leads is therefore about 0.04 $\mu_B$ ($N=4$) 
and 0.06 $\mu_B$ ($N=5$), still a very small fraction of the total moment. 
The Pt atoms in the nanowire clearly drive the magnetization, but at least in this geometry 
a giant moment\cite{shen69} apparently does not form.

We also obtained the total energies $\Delta$ of various metastable states above 
the parallel magnetic ground state. The transversely magnetized state has a much 
higher energy than the ground state, about $\Delta_\perp=50$~meV and $\Delta_\perp=80$~meV 
for the four-atom and five-atom wires, respectively. The three-atom contact
does not have a transversely magnetized state, but there is still 
a battier of $\Delta_{NM}=12$ meV between the two equivalent parallel magnetization
ground states. 
We did not repeat the nanocontact calculations for general chain interatomic 
distances $d$. The qualitative result can still be roughly estimated by simply rescaling 
infinite wire ground state energies of Ref.~\onlinecite{colossal} and assuming a total 
of $N-2$ equivalent magnetic Pt atoms in the nanocontact. In that way we 
anticipate that for example at  $d= 2.5$~\AA, for 
all $N=3$--5 there should be a (parallel) magnetized ground state but no
transversely magnetized state whatsoever (colossal anisotropy) with an
estimated barrier between the two parallel polarizations about 10-30 meV.
These large magnetic anisotropy barriers should effectively impede the 
rotation and/or the flip of the nanocontact magnetization as a whole. A high 
barrier should also to some extent hamper fluctuations. While a detailed 
appraisal and treatment of fluctuation effects is beyond the scopes 
of this work, we can still use this result for some tentative conclusions. 
As a consequence of the high barrier, a full magnetization reversal due to 
occasional transverse anisotropies and/or thermal fluctuations should be rare. 
Local spin fluctuations, which renormalize but do not reverse magnetization 
become competitive with the global ones. Interestingly, the Zeeman coupling 
energy $\mu_B B M_{S}^{tot}$ of an external field $B$ may only match the magnitude 
of the anisotropy barrier at field values as large as 100 Tesla. Therefore the 
anticipated effect of a field of ordinary magnitude is negligible.

We tentatively conclude that the Pt nanowire contact is in effect 
an Ising-like nanomagnet with a large blocking temperature ($k_BT\sim$ 10-50 meV), and 
thus with utterly infrequent thermal magnetization flips at criogenic 
temperatures.  This assessment will need a revision at reduced or zero strains,
where the chain interatomic distances may approach their minimum value
near $d_0=2.35$ \AA~and nanowire magnetism, though still finite, becomes
weaker. The nanocontact in this regime should still exibit the property of colossal 
anisotropy the possible consequences of which will be examined elsewhere.

\section{Ballistic conductance}

The main measurable physical quantity in a nanocontact is its 
electrical conductance. We calculated the ballistic electron conductance
within the Landauer-B\"uttiker linear response formalism, appropriate as was
discussed above when all the system parameters, including magnetization, 
can be treated as static. At sufficiently small voltage, ballistic conductance  
$G$ is proportional to the total electron transmission at the Fermi energy, 
$G=(1/2)G_0 T(E_F)$. In order 
to calculate the nanocontact transmission we considered 
the geometry shown in Fig.~\ref{mag_3a4a5a} as the nanocontact scattering 
region, to be joined ideally to semi-infinite bulk Pt leads on both sides. 
The scattering problem between incoming and outgoing Bloch waves 
is then solved using the complex band method (for details 
see Ref. \onlinecite{choi, our_method}) whereby the transmission matrix and 
the total transmission is subsequently computed.
For our plane-wave based FR calculations we use a recent extension 
of a method\cite{cond_soc} accounting for SOC effects, and including
proper treatment of two-component spinor Bloch wave functions. Since the system 
has the supercell (artificial) 2D periodicity in the $xy$ plane, perpendicular
to the transport direction, we averaged 
the transmission over the corresponding 2D Brillouin Zone (BZ) using 
21 {\bf k} points in the irreducible 
part of the BZ. This level of sampling was found to be needed and sufficient
to get rid of spurious oscillations in transmission function and 
yields our best approximation to electron transmission 
in the true isolated nanocontact.

\begin{figure}
\hskip-3mm
\includegraphics[width=8.5cm]{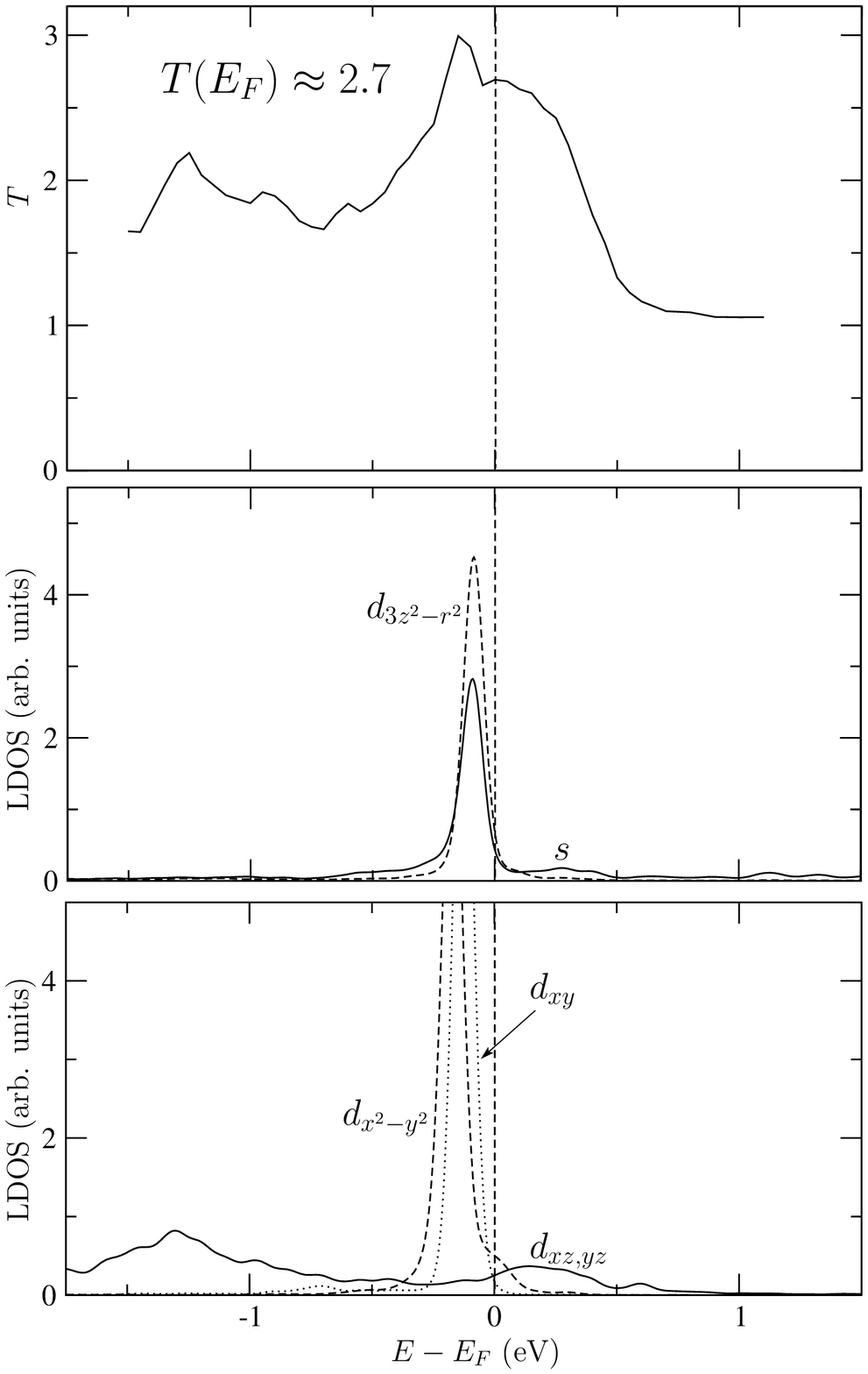}
\caption{\label{t_sr_3a}
Transmission coefficient (per spin) versus energy for the three-atom nanowire Pt contact at $d=2.66$ \AA~(upper panel) in the SR approximation, where spin orbit coupling is not included, and the contact is nonmagnetic.
Note the anticipated value of ballistic conductance of 2.7 $G_0$, considerably higher
than typical experimental values in the range $1.5-2.0~G_0$. Lower panels show the LDOS projected on different atomic orbitals of the middle nanowire atom.}
\end{figure}

We started with the SR calculation for the three-atom nanowire contact 
and present in Fig.~\ref{t_sr_3a} the transmission function (per spin) 
versus the energy. In this approximation the Pt nanocontact is nonmagnetic.
In order to understand the various features in 
the transmission curve we plot on the lower panels the LDOS projected 
on different atomic orbitals of the middle nanowire atom. At 
energies $E-E_F>0.5$ eV the transmission is very close to $1$. 
In this energy region in the infinite wire there is only one $s-p_z$-like 
band for each spin (see upper panel of Fig.~1). This is very
broad band, high kinetic energy states which generally exhibit very 
little reflection by obstacles such as the nanocontact, yielding
nearly free propagation and transmission close to unity. At lower energies the
$d$ states make their appearance and the transmission starts growing significantly. 
The main contribution here comes from $d_{xz},d_{yz}$ states which 
form quite broad $m=\pm 1$ bands in the infinite wire. Just below 
the Fermi energy one can see several sharp features in the LDOS 
both for $s,d_{3z^2-r^2}$ and for $d_{xy},d_{x^2-y^2}$ orbitals 
which are related to narrow nanowire bands of $m=0$ and of $m=\pm 2$ 
symmetry, respectively, lying close to the Fermi energy. These LDOS
peaks give rise to the sharp feature in the transmission 
function at the energy $E-E_F\sim -0.2$ eV.
From the transmission at the Fermi energy we obtain 
the ballistic conductance of about $2.7~G_0$. The transmission curve 
shown in Fig.~\ref{t_sr_3a} looks quite similar 
to that presented recently by Fern\'andez-Rossier {\it et al.}\cite{fernandez} who 
with a slightly different geometry and larger interatomic distances
in the nanowire found the conductance of about $2.3~G_0$. Our 
conductance is also somewhat higher than the values of $\sim2~G_0$ 
calculated for straight monatomic Pt wires in Refs.~\onlinecite{garcia-suarez, strange}. 
Break junction experiments generally report a broad peak 
in conductance histogram centered at values between $1.5~G_0$ and 
$2~G_0$.\cite{smit1,krans,yanson1,nielsen,smit2,untiedt} 
Recently, Nielsen and co-workers\cite{nielsen} argued that even lower conductances 
(down to $1.5~G_0$) should be assigned to monatomic wires while single-atom 
contacts should have conductances close to $2~G_0$.

\begin{figure}
\hskip-3mm
\includegraphics[width=8.5cm]{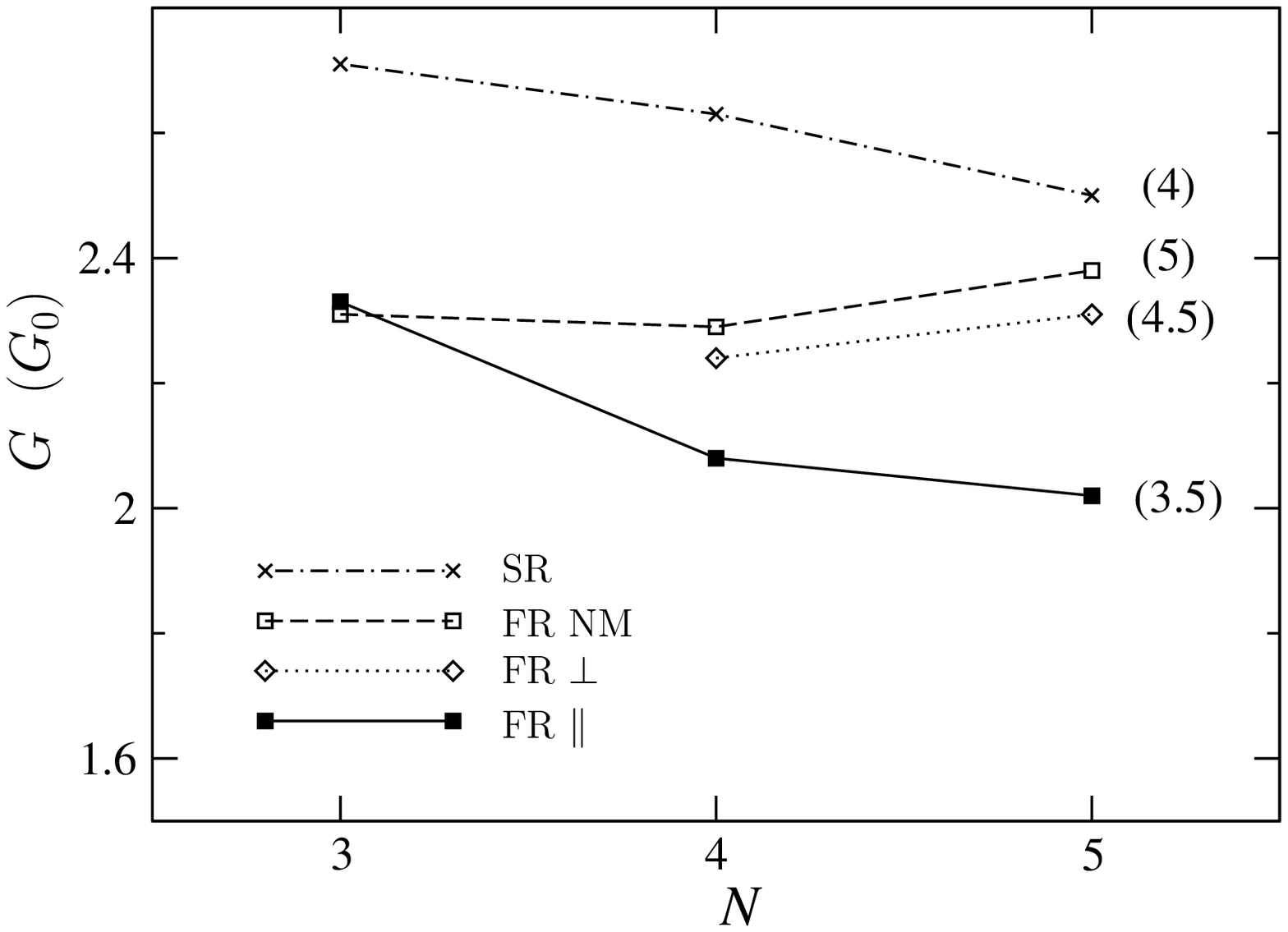}
\caption{\label{g_3a4a5a}
Ballistic conductance of three-, four-, and five-atoms nanowire Pt contacts.
Here SR and FR calculations are compared including in the latter case,
the conductances for both parallel and transverse magnetic states as well as for the nonmagnetic state. 
The FR state with parallel magnetization corresponds to the nanocontact ground state,
with a ballistic conductance of about $2~G_0$.  
The ideal contactless maximal conductance of a perfect 
infinite nanowire (deduced from the channel number in the band structure of Fig. 1) is also provided for each case. 
}
\end{figure}

We now turn to FR ballistic conductance calculations, our best approximation. 
We collect and compare in Fig.~\ref{g_3a4a5a} 
the SR and FR results for nanowire contacts of different lengths $N=3$, 4, 5. 
On the right side of the figure we also mark the nominal maximum ``conductances'' for the ideal 
infinite monatomic wire for each case, $G=(N_{chan}/2)G_0$ (see Fig.~\ref{bands}). 
As a general effect, we see that the conductance is lowered when SOC is
included. For the three-atom nanocontact the FR conductance is about $2.3~G_0$ 
for both magnetic and nonmagnetic states, about 15\% lower
than the SR conductance ($\sim2.7~G_0$). 
This result could not have been anticipated from a simple consideration of 
the infinite nanowire bands of Fig.~\ref{bands} -- one could rather have expected 
the highest conductance for the FR nonmagnetic state ($N_{chan}=10$)
relative to the SR case ($N_{chan}=8$), and finally a much lower conductance for the
magnetic state ($N_{chan}=7$). Such differences originate mainly from the 
shortness of the nanocontact -- a three-atom nanowire is far from infinite.
Magnetism is significantly weakened (see upper panel of Fig.~\ref{mag_3a4a5a}), 
and, as a consequence, so are the effect of magnetism on conductance. 

When the nanowire becomes longer the magnetism in 
the nanowire gets stronger. Conductance through the magnetic nanocontact 
drops down to $\sim 2~G_0$ and differs more and more 
from that of the nonmagnetic state. For $N=5$ this difference
is as large as $0.4~G_0$. 
We note that recently Fern\'andez-Rossier and co-workers\cite{fernandez} 
performed SR calculations (where spin-orbit effects were disregarded) 
and also reported the formation of magnetic moment in Pt nanowires under 
some significant strain. They also found that the conductance is lower for 
the magnetic state, in agreement with our calculations.

For nanocontacts with $N=$ 4 and 5, we calculate and predict
anisotropic magnetoresistance, namely the state with transverse magnetization 
has a higher conductance than the (ground) state with 
parallel magnetization. Our calculations predict an AMR of $\sim 8$\% and $\sim 15$\% 
for $N=$ 4 and 5, respectively. Note that these values for 
AMR are still noticeably smaller than the ideal value given earlier for an infinite nanowire
($\sim 28{\%}$). We underline again however that due to the high barrier 
it should be generally very hard if not impossible to turn the 
magnetization away from the parallel direction, and measure directly 
the AMR of a Pt nanocontact with practically attainable fields. 

Looking at conductances of various states of the five-atom Pt nanocontact 
one can see that, unlike the case of a shorter three-atom wire, they are 
now arranged almost in the same order as those of the ideal infinite wire.  
The only remaining disagreement is for 
the SR conductance -- it is still higher than the conductance of, e.g., FR 
nonmagnetic state. This can be rationalized by looking at the bands of
Fig.~\ref{bands} and noticing that even though the number of conductance 
channels is larger for the FR nonmagnetic state than for the SR case (10 against 8) 
some of the bands touch the Fermi level close to their very edges. Here 
the electron group velocity is small, and so is the contribution to conductance.
 
\section{Discussion, and spin filtering properties}

The above density functional calculations show that short but well defined 
monatomic nanowire segments forming at Pt nanocontacts should be spontaneously  
magnetic. The overall magnetic energy gain increases with the number of atoms
in the wire and with the strain. It is quite large, reaching up to 90 meV for $N=5$ and 
large strains. For the chosen nanowire interatomic distance $d=2.66$~\AA,
the magnetic anisotropy is also exceedingly large -- ``colossal" for $N=3$ 
and ``giant'' for  $N=4$, 5, the easy magnetization axis parallel to the nanowire. 
Ballistic conductance is sensitive to the onset of local magnetism, whose 
presence causes a drop of the order of 10\% relative to the nonmagnetic state
(a state which however is unstable). 

The question is, whether and how it will be possible to obtain a direct 
experimental evidence of the presence of nanocontact magnetism in Pt, 
a task which is not easy at the present stage. The most common type of 
evidence of magnetism at quantum dot and molecular nanocontacts 
is a Kondo zero bias anomaly.\cite{kondo} We note here that our system is not a regular 
Kondo system, on account of giant anisotropies and of other elements including 
proximity to ferromagnetism in the leads, and will defer this aspect for the 
time being, and concentrate on bias voltages slightly away from zero, where
these effects should be irrelevant. Here, assuming the time of traversal 
of a ballistic electron is short enough, the nanocontact should effectively
exhibit a static or slowly varying magnetic moment, with some interesting
even if speculative consequences. 

\begin{figure}
\hskip-3mm
\includegraphics[width=8.2cm]{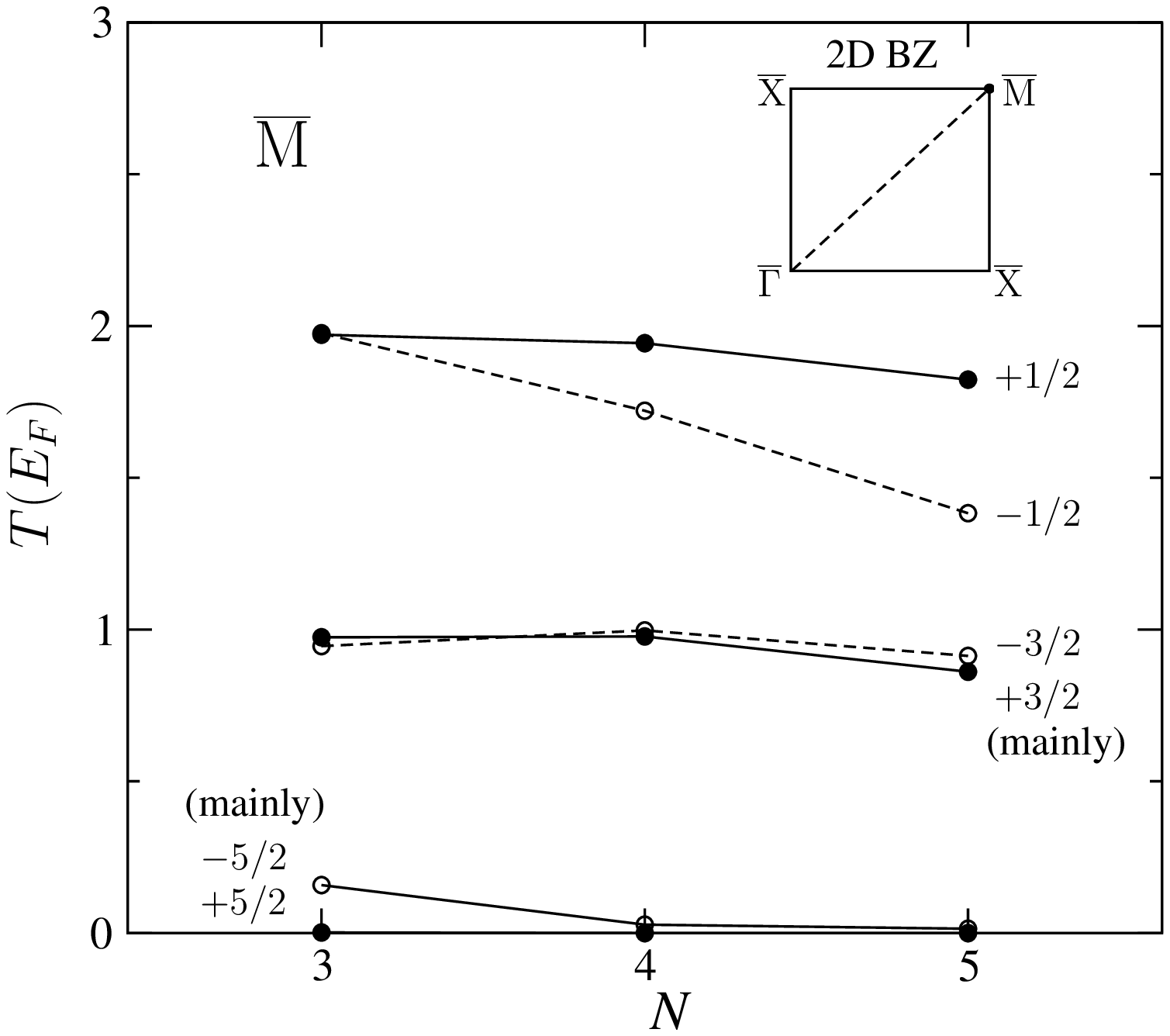}
\caption{\label{g_mj}
Eigenchannel decomposition of the FR total transmission at the Fermi
energy calculated at the $\overline{\rm M}$ 2D {\bf k} point (see the inset). 
Here (as well as at the $\overline{\Gamma}$ point), due to the symmetry,  
the half-integer $m_j$ can be assigned to each 
transmission eigenchannel (see text).
}
\end{figure}

One of them is a spin filtering effect. In ferromagnetic nanocontacts (such as Ni or Co) 
majority and minority spin channels have very different conductances, the former smaller that 
the latter, which leads to a spin-polarized current in the steady state.
The effect of that may be difficult to detect on account of the large pre-existing
spin polarization of the leads. In Pt nanocontacts, the advantage is that
the leads are not magnetic, and therefore any evidence of magnetism can
be safely attributed to the nanocontact.
In order to illustrate the possibility of a spin polarized net current 
for our {\it locally} magnetized Pt nanocontacts we show in Fig.~\ref{g_mj} 
the eigenchannel decomposition of the total FR transmission
at the Fermi energy as a function of $N$ calculated at the $\overline{\rm M}$ point of the
2D BZ. We label all the eigenchannels by half-integer $m_j$ as follows.
Since the nanocontact chain is magnetic in parallel direction, the symmetry group
of the $\overline{\rm M}$ point (and also of the $\overline{\Gamma}$ point) is
the double group of $C_{4}$, the group of our slab. All transmission eigenchannels 
can be classified according to its four one-dimensional irreducible representations, 
in turn expressible as linear combinations of representations of the larger infinite nanowire group
$C_\infty$, labeled by all half-integer $m_j$. In this way, $m_j$ and $m_j+4n$ (with $n$ integer) can mix to form the same representation of $C_{4}$ double group and the 
corresponding states will generally get mixed in the presence of square symmetry contacts. 
This is the case, for example, for $+3/2$ and $-5/2$ as well as for $-3/2$ and $+5/2$ states.

We note that, as illustrated in Fig.~\ref{g_mj} for the $\overline{\rm M}$ $\bf k$ point, 
magnetism causes in general a conductance imbalance between $+m_j$ and $-m_j$ channels
(here, mainly between $\pm 1/2$ channels) which becomes larger for longer wires due to 
stronger magnetism. Therefore, as current  flows through the Pt nanocontact, there 
will be an accompanying magnetization (spin, and orbital) flow as well. During the 
time $\tau$ (presently unknown, but assumed long) while the nanomagnet magnetization 
direction does not flip, this should lead to a magnetization
accumulation uphill, and a corresponding depletion downhill. The amount of accumulation is 
determined by the actual conductance filtering asymmetry, which we calculate by the current, which 
can be controlled, and by the rate of magnetization decay in the bulk Pt leads, which is 
uncontrolled but probably high on account of a large DOS at the Fermi level and large SOC.
The actual calculated amount of selective filtering shown in Fig.~\ref{g_mj}
is not large, but even if modest, it could be important, because the Pt
leads are themselves nonmagnetic, and all magnetic effects can only be attributed
to the nanocontact itself. In fact, current in presence of local magnetism at the 
Pt nanocontact should alter the size and extent of the magnetization tails into 
the bulk-like Pt leads. In particular, a ferromagnetic giant tail moment, absent
in equilibrium, might be expected to form uphill of a nanocontact, previously prepared
in a well defined polarization state by e.g., field cooling. In these conditions,
a spontaneous reversal of magnetization would imply a large transient magnetic
reorganization, the giant tail shifting from uphill to dowhill. This could in turn 
reflect in an observable transient effect on conductance. However, the magnitude
of this effect remains at this stage unpredictable.

 \section{Conclusions}
We carried out fully-relativistic density functional study of electronic, magnetic, 
and transport properties of nanowire Pt contacts. We studied model nanowire contacts 
with $N=3$, 4, and 5 atoms in the wire and found that they remain locally magnetic 
despite the presence of nonmagnetic Pt leads, and that magnetism is stronger for longer wires
and for larger strain.
Spin-orbit effects are crucial to the phenomenon and give rise to a 
very large spin anisotropy for $N=$ 4 and 5 with easy axis along the nanowire. 
The energy of a transversely magnetized state
is much higher, roughy $\sim50$~meV and $\sim80$~meV for $N=4$ and 5, respectively. 
For the shorter nanocontact with $N=3$ there is no transversely magnetized state, and
the anisotropy is ``colossal''. These large barriers should provide a high blocking 
temperature below which thermal fluctuations are ineffective, and nanowire magnetism could 
be observed. Kondo phenomena are not addressed here, but are expected to be
strongly influenced by anisotropy, and by proximity to ferromagnetism in Pt.

From our calculated Landauer-B\"uttiker ballistic electron transport, we conclude that 
inclusion of both SOC and magnetism is important and lowers the ballistic conductance 
by about $15\div20$\%. The lowest conductance, $G\sim2~G_0$, occurs in the ground state 
with magnetization parallel to the wire. These values are close to those reported in 
break junctions experiments ($G\sim 1.5-2.0~G_0$). A conductance of 1.5 $G_0$ is 
however below our calculated value. During the breaking process, the symmetry 
of the nanocontact is expected in general to be lower than that implied by our assumed geometry.
A lower symmetry might act to block some channels or anyway decrease their 
transmissions lowering even more the total conductance. 
For example, it has been recently shown that zigzag Pt nanowires have a lower 
conductances ($\sim 1.5~G_0$) with respect to the straight ones ($\sim 2~G_0$).
While we note that a zigzag configuration is expected to be removed by stress,
this example does make the point. 

For long nanowire contacts ($N=4$, 5) the conductance was found to 
increase when the magnetization is rotated and becomes perpendicular to the wire 
axis, a potential effect of anisotropic magnetoresistance (AMR). We calculated 
the AMR magnitude of about 8\% and 15\% for $N=4$ and 5, respectively. 
The AMR could be in principle observed by applying a field to turn the magnetization
perpendicular to the nanowire. However, due to the large magnetic anisotropy barrier 
the static deviation of the magnetization from the easy axis would be generally
small for magnetic fields attainable in laboratory conditions. Quantum tunneling 
of magnetization,\cite{chudnovsky1988} although not discussed here and in principle 
not ruled out, should be strongly hindered by the large or colossal axial anisotropy, 
by temperature, and possibly by additional orthogonality phenomena due to 
the presence of the Fermi sea. 

Our results suggest some qualitative hints for the possible observation of
our predicted magnetic nanocontact phenomena, which at the moment remain exquisitely
theoretical. It might still be possible to observe the effects of thermal or quantum flipping 
of magnetization in the form of noise in the current. Spin filtering effects could 
also be detectable. Finally, Kondo zero bias anomalies, although not addressed here, could 
provide a ``smoking gun" for Pt nanowire magnetism.

\section{Acknowledgments}

We are grateful for discussions with C. Untiedt. Work sponsored by MIUR PRIN/Cofin 
Contract No. 2006022847 as well as by INFM/CNR ``Iniziativa transversale calcolo parallelo". 
Part of the calculations have been done on the SP5 at CINECA in Bologna.

\end{document}